\documentclass[a4paper,11pt]{article}
\usepackage{jcappub} 
\usepackage{lineno}
\arxivnumber{\textcolor{red}{2502.13914}} 
\title{Thermodynamic Topological Classes of Family of Black Holes with NUT-Charge}

\author[a]{Mustapha Azreg-A\"{\i}nou,}
\author[b]{Muhammad Rizwan}
\affiliation[a]{Ba\c{s}kent University, Engineering Faculty, Ba\u{g}l{\i}ca Campus, Ankara, Turkey}
\affiliation[b]{Department of Mathematics, Faculty of Engineering and Computing, National University of Modern Languages, 
H-9, 43000, Islamabad, Pakistan}
\emailAdd{azreg@baskent.edu.tr}

\emailAdd{mrizwan@numl.edu.pk}

\abstract{We introduce new analytical methods for treating black-hole thermodynamic topology, and thus depart from the usual treatment known in the literature. As application we investigate the universal thermodynamic topological classes of a family of black holes with NUT-charge. A charge-dependent thermodynamic topological transition from one class to another class is discussed.}

\begin{document}
\maketitle

\flushbottom

\section{Introduction}\label{sec:intro}
Black holes, among the most mysterious objects in nature, have gained significant attention due to recent observational breakthroughs, such as the direct imaging of black hole shadows \cite{vagnozzi2023horizon, grenzebach2014photon} by the Event Horizon Telescope (EHT) \cite{collaboration2019first, akiyama2022first} and the detection of gravitational waves from black hole mergers by LIGO-VIRGO \cite{abbott2016observation,abbott2016gw151226}. These compact yet massive objects are at the forefront of theoretical advancements, offering insight into the fundamental nature of quantum gravity.  Black hole thermodynamics \cite{hawking1974black, hawking1975particle}, along with its extended version \cite{kubizvnak2017black} which considers the cosmological constant as a thermodynamic parameter, has led to numerous phase transitions and phase structures \cite{kastor2009enthalpy,kubizvnak2012p,azreg2015}, particularly the Hawking-Page phase transition \cite{hawking1983thermodynamics}  holds significant importance for gauge theory through the AdS/CFT correspondence.

The universal thermodynamic classification of black holes has recently been proposed, categorizing all black hole solutions into three classes \cite{wei2022black, wu2023topological}, which were later extended to four topological classes \cite{wei2024universal,zhu2025universal,rizwan2025universal,wu2023topologicalAds}. This classification is based on the concept that black hole solutions can be viewed as topological defects within the thermodynamic parameter space, and each black hole state relies on topological numbers. The topological classes of various black hole spacetimes have been studied extensively, with some of these studies being referenced in \cite{bai2023topology, liu2023topological,fang2023revisiting, rizwan2023topological,alipour2023topological,Gashti1,Gashti2}. The topological classification of the neutral Lorentzian NUT-charged spacetime has been investigated \cite{wu2023classifying}, while the classification of the Lorentzian-charged Taub-NUT spacetime has been addressed \cite{wu2023consistent}. 

In the scientific literature, the thermodynamic classification of black holes has relied on the existence of a manifold $\mathcal{M}$ with boundary $\partial\mathcal{M}$ and of a vector field $\phi$ supposed to be perpendicular to $\partial\mathcal{M}$. In order to satisfy the requirement that the unit vector $\phi/||\phi||$ be perpendicular to $\partial\mathcal{M}$, this has laid to introduce a vector field with a component that is not analytic on the boundary. This construction does not fulfill the requirement of Poincar\'e-Hopf Index Theorem for manifolds with boundary~\cite{book3}. In this work, taking advantage of the Poincar\'e-Hopf Index Theorem for manifolds without boundary, we introduce new analytical methods that allow for the topological selections and classifications of vector fields. In this approach, the vector field $\phi$ is analytic everywhere on the manifold. We first introduce a method for classification of vector fields based on the difference of the number of maxima and minima of the potential function that generates the vector field. We then specialize to topological thermodynamics, re-derive the four cases of thermodynamic topological classes, and apply them to investigate the thermodynamic topological classification of a set of black hole solutions.

In the thermodynamic topological classification of black hole solutions, the vector field $\phi$ is defined, as we mentioned, as the gradient of some function $\tilde{F}(r_h,\Theta)$ which is the off-shell free energy $F(r_h)$ extended with an additional parameter $\Theta\in (0,\pi)$ as $G(\Theta)=\csc\Theta$, that is, $\tilde{F}(r_h,\Theta)=F(r_h)+G(\Theta)$ . Applying the Duan mapping current theory \cite{duan20182, duan1998bifurcation} each zero point of this vector field can be related to a black hole state \cite{wei2022black}. Our proposed method is based on Poincar\'e-Hopf Index Theorem for manifold without boundary.

The paper is organized as follows. In Sec.~\ref{secT} we introduce relevant notions on vector fields in topology along with the Poincar\'e-Hopf Index Theorem and Euler characteristic. One of the components of the vector field no longer includes the $\csc$ function, which is replaced by the $\sin$ or $\cos$ function. Since the manifold has no boundary, a variety of selections of the trigonometric component of the vector field is offered. In Sec.~\ref{secTC} we discuss topological classes of general vector fields and determine a relation between the topological charge and the relative number of extrema of the potential function. In Sec.~\ref{secTTC} we specialize to thermodynamic topological classes. In Sec.~\ref{secAp} we investigate the thermodynamic topological classification of a set of black hole solutions and we discuss a charge-dependent thermodynamic topological transition from one class to another class. We summarize our results in Sec.~\ref{secConc}. As a by-product we derive useful formulas for Reissner-Nordstr\"{o}m-NUT-AdS black hole. An appendix section is added for the evaluation of the Euler characteristic of a cylinder.

\section{Preliminaries: Topology\label{secT}}
In this section, we define some notions of differential topology. Let $\mathcal{M}$ be a smooth oriented surface and $\phi$ be a smooth varying vector field on it. For any zero $X\in \mathcal{M}$ of $\phi$ [simple or multiple, also called a critical point (CP)], that is, $\phi (X) = 0$, and for any $\epsilon >0$ we define the disk $\mathcal{D}(X,\epsilon)$ of center $X$ and radius $\epsilon$. If for any $Y\in \mathcal{D}(X,\epsilon)\setminus\{X\}$ we have $\phi(Y)\neq 0$, we say that $X$ is an isolated zero of $\phi$.

In the subsequent sections, we will consider vector fields $\phi$ that are gradients of some function $\tilde{\mathcal{F}}(r_h,\Theta)$ where $(r_h,\,\Theta)$ is a coordinate chart on $\mathcal{M}$. The change in angle that the unit vector $\phi /||\phi||$ makes after traveling counterclockwise once around any small closed path that surrounds the point $X=(r_h=x,\,\Theta=\Theta_0)$ with $\phi (X) = 0$, or simply a circle centered at $X$, divided by $2\pi$ is called the index (or winding number) of $\phi$ at the isolated zero $X$ and is denoted by $\text{ind}(\phi,X)\in \mathbb{Z}$, which is a positive or negative integer. The index does not depend on the choice of the local coordinates $(r_h,\,\Theta)$ or on the shape of the small closed path enclosing $X$. By the Poincar\'e-Hopf index theorem, the sum of all indices of $\phi$ is an invariant of $\mathcal{M}$~\cite{Poincare1,Poincare2,Hopf,book}. The theorem has two versions, the one concerning manifolds with boundary and the other applies to manifolds without boundary. In this work, we are concerned with the latter version, which states\vskip8pt

\noindent\textbf{Poincar\'e-Hopf Index Theorem for manifolds without boundary:}
Let $\phi$ be a vector field with only isolated CPs on a surface $\mathcal{M}$ that can be orientable or nonorientable, and let $X_1$, $X_2$, ..., $X_i$ be the set of all isolated CPs (zeros) of $\phi$. Then $\sum _{i}\text{ind}(\phi,X_i)=\chi(\mathcal{M})$, where $\chi(\mathcal{M})$ is the Euler characteristic of $\mathcal{M}$.  \vskip8pt

The theorem relates the Euler characteristic, which is a global and topological quantity, to the indices of CPs, which are local and analytical quantities.

Since our vector field is a gradient of some function, the zeros of $\phi$ correspond to the extreme values of $\tilde{\mathcal{F}}$. As we shall see, when these extrema are local maxima or minima, the index is positive, and if the extrema are saddle points, the index is negative.

\begin{figure}[!htb]
	\centering
	\includegraphics[width=0.4963\textwidth]{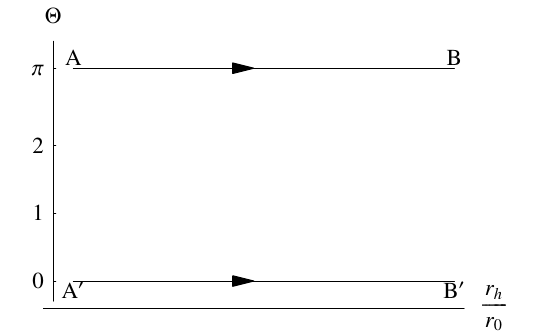}
	\includegraphics[width=0.4963\textwidth]{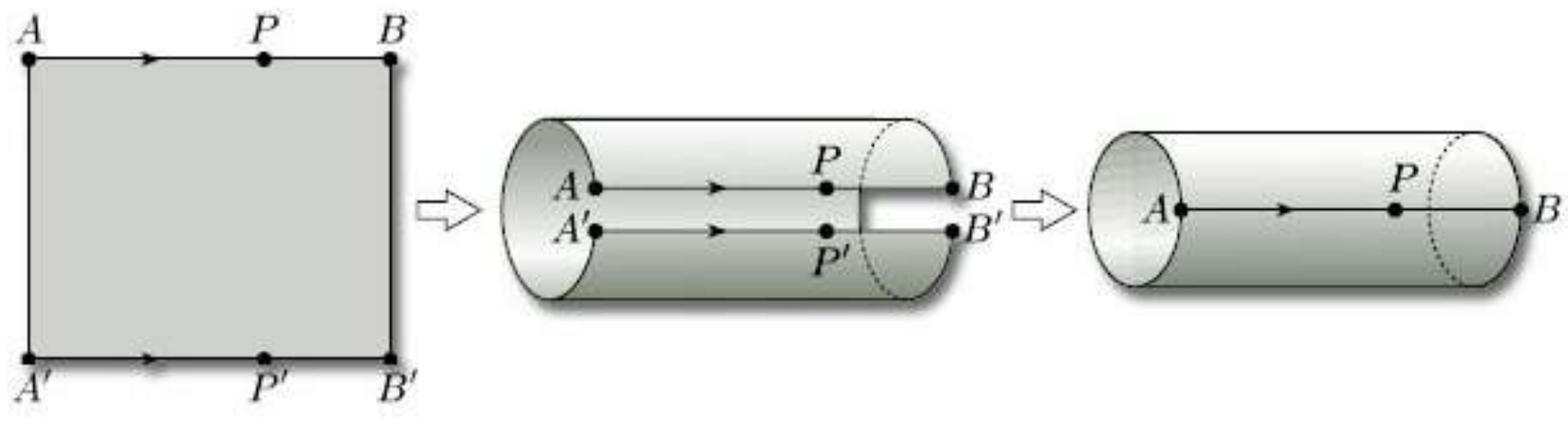}
	\caption{Representation of the manifold $\mathcal{M}$ as a cylinder upon identifying and gluing the lines $\Theta=\pi$ and $\Theta=0$. Here $r_0$ is a scale parameter defined in the caption of Fig.~\ref{FigTNUT} and in the text.}
	\label{Figid}
\end{figure}

Our function $\tilde{\mathcal{F}}(r_h,\Theta)$ will be the off-shell free energy $F(r_h)$, which depends on the horizon location $r_h$ augmented by some function $G(\Theta)$ of $\Theta$
\begin{equation}\label{F0}
	\tilde{\mathcal{F}}(r_h,\Theta)=F(r_h)	+ G(\Theta)\,.
\end{equation}
For the chart of our manifold $\chi(\mathcal{M})$, $r_h$ runs from $0\text{ (or }r_m)\to\infty$ and we \emph{want} that $\Theta$ runs from $0 \to\pi$. Since we want our manifold to be without boundary, this yields the representation shown in Fig.~\ref{Figid}, where we identify the horizontal line $\Theta=\pi$ with the horizontal line $\Theta=0$ to obtain a cylinder. This identification ensures that the vector field $\phi$ is smooth as one crosses the line $AB$ with a continuous derivative $\partial_{\Theta}\phi^\Theta$. As is well known, for a cylinder $\chi(\mathcal{M})=0$ (see the appendix for a derivation of $\chi(\mathcal{M})$ by subdivision, and see~\cite{book2} for a derivation by triangulation).

Since the vector field $\phi$ must be smooth on the cylinder $\mathcal{M}$, the choice usually made in the literature, of $G(\Theta)=\csc\Theta$, is no longer suitable. In our case we take $G(\Theta)= (1/2)\sin 2\Theta$ or $G(\Theta)= (1/2)\cos 2\Theta$
\begin{equation}\label{F1}
	G_1(\Theta)=\frac{\sin 2\Theta}{2}\quad\text{ or }\quad G_2(\Theta)=\frac{\cos 2\Theta}{2}
\end{equation}
The vector field $\phi$ has thus the general components
\begin{equation}\label{F2}
	\phi^{r_h}=\partial_{r_h}F(r_h), \; \phi^{\Theta}=\cos 2\Theta \quad\text{ or }\quad\phi^{r_h}=\partial_{r_h}F(r_h), \; \phi^{\Theta}=-\sin 2\Theta\,.
\end{equation}

\begin{figure}[!htb]
	\centering
	\includegraphics[width=0.4963\textwidth]{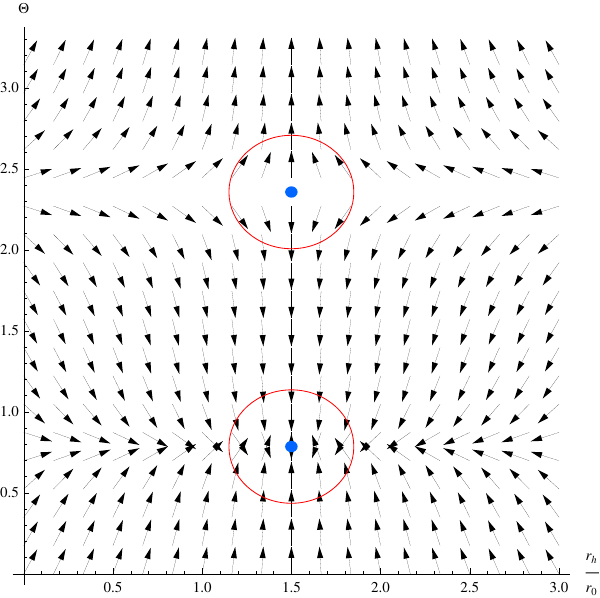}
	\includegraphics[width=0.4963\textwidth]{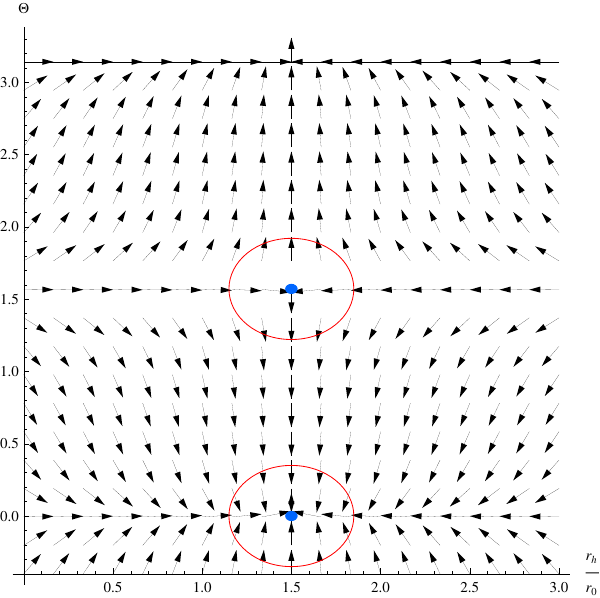}
	\caption{Vector field plots of the unit vector field $\phi/||\phi||$ for the Lorentzian Taub-NUT black hole~\eqref{LTNUT} taking $\tau=6\pi r_0$~\eqref{LTNUT2}. Left Panel: $G(\Theta)= G_1(\Theta)=(1/2)\sin 2\Theta$. Right Panel: $G(\Theta)= G_2(\Theta)=(1/2)\cos 2\Theta$. In both panels the blue dots represent CPs located at $(r/r_0,\Theta)=(3/2,\pi/4),\,(3/2,3\pi/4)$ for the left panel, and at $(r/r_0,\Theta)=(3/2,0),\,(3/2,\pi/2)$ for the right panel. For the upper CP in each panel, the change in the angle that the unit vector $\phi /||\phi||$ makes after traveling counterclockwise once around the red closed curve is $-2\pi$ so that the index or the winding number is $-1$. And for the lower CP in each panel, the change in the angle that the unit vector $\phi /||\phi||$ makes after traveling counterclockwise once around the red closed curve is $+2\pi$ so that the index or the winding number is $+1$. Here, $r_0$ represents an arbitrary length scale defined by the size of a cavity surrounding the black hole~\cite{wei2022black}.}
	\label{FigTNUT}
\end{figure}

The function $G_1(\Theta)$~\eqref{F1} has a maximum at $\Theta=\pi/4$ and a minimum at $\Theta=3\pi/4$. If $F(r_h)$ has a maximum value at some zero $r_h=x$ of $\phi^{r_h}=0$, in this case $\tilde{\mathcal{F}}(r_h,\Theta)$ will have a maximum at the zero ($r_h=x,\,\Theta=\pi/4$) of the vector filed and the corresponding index will be positive. In this case too, $\tilde{\mathcal{F}}(r_h,\Theta)$ will have a saddle point at the other zero ($r_h=x,\,\Theta=3\pi/4$) of the vector filed and the corresponding index will be negative, so that the sum of the indices is $\chi(\mathcal{M})=0$ by the Poincar\'e-Hopf index theorem. It is easy to see that if now $F(r_h)$ has a minimum value at some zero $r_h=x$ of $\phi^{r_h}=0$, then the index of the zero ($r_h=x,\,\Theta=\pi/4$) will be negative and the index of the other zero ($r_h=x,\,\Theta=3\pi/4$) will be positive, so that the sum of the indices is again 0. Illustrations are shown in Fig.~\ref{FigTNUT}, which is a generic figure, with $r_0$ being an arbitrary length scale defined by the size of a cavity surrounding the black hole~\cite{wei2022black}. The upper blue point in the left panel is the CP corresponding to $\Theta=3\pi/4$ and has a negative index, and the lower blue dot is the CP corresponding to $\Theta=\pi/4$ and has a positive index. As to the value of the index, it is easy to show that this is either $+1$ or is $-1$, because $\cos 2\Theta$ vanishes only 2 times as one moves on the circle surrounding the CP. In Fig.~\ref{Figmax}, we show how the index is evaluated in case $F(r_h)$ has a maximum value at the CP ($r_h=x,\,\Theta=3\pi/4$).

The function $G_2(\Theta)$~\eqref{F1} has a maximum at $\Theta=0$ and a minimum at $\Theta=\pi/2$. So, the whole picture is similar to the previous case where we have to replace the values $\pi/4$ and $3\pi/4$ by $0$ and $\pi/2$, respectively. In this case the CPs are ($r_h=x,\,\Theta=0$) and ($r_h=x,\,\Theta=\pi/2$). The illustrations are shown in the right panel of Fig.~\ref{FigTNUT}. The upper blue point is the CP corresponding to $\Theta=\pi/2$ and has a negative index, and the lower blue dot is the CP corresponding to $\Theta=0$ and has a positive index. A similar argument to that given in the previous paragraph allows one to conclude that the value of the index is $+1$ or $-1$.

Based on the explanations given in the last two paragraphs, it is a mere matter of choice to use either function $G_1(\Theta)$ or $G_2(\Theta)$ for the determination of the index and for further investigations.

For the remaining parts of this work, we will not sketch plots similar to the vector field plots of Fig.~\ref{FigTNUT} for each black hole solution; rather, we will determine the \emph{nature} of the extreme values of the function $F(r_h)$. The rules are as follows.
\begin{align}
	&\text{For }G_1(\Theta)=\frac{\sin 2\Theta}{2}\,,\nonumber\\
	&\text{If }F(r_h=x)=\text{ max},\quad X_1=(x,\,\pi/4),\quad  X_2=(x,\,3\pi/4)\Longrightarrow \nonumber\\
	\label{r1}&\text{ind}(\phi,X_1)>0 \text{ and }\text{ind}(\phi,X_2)<0\,,\\
	&\text{If }F(r_h=x)=\text{ min},\quad X_1=(x,\,\pi/4),\quad  X_2=(x,\,3\pi/4)\Longrightarrow \nonumber\\
	\label{r2}&\text{ind}(\phi,X_1)<0 \text{ and }\text{ind}(\phi,X_2)>0\,.
\end{align}
\begin{align}
	&\text{For }G_2(\Theta)=\frac{\cos 2\Theta}{2}\,,\nonumber\\
	&\text{If }F(r_h=x)=\text{ max},\quad X_1=(x,\,0),\quad  X_2=(x,\,\pi/2)\Longrightarrow \nonumber\\
	&\text{ind}(\phi,X_1)>0 \text{ and }\text{ind}(\phi,X_2)<0\,,\\
	&\text{If }F(r_h=x)=\text{ min},\quad X_1=(x,\,0),\quad  X_2=(x,\,\pi/2)\Longrightarrow \nonumber\\
	&\text{ind}(\phi,X_1)<0 \text{ and }\text{ind}(\phi,X_2)>0\,.
\end{align}

An important generalization is in order. Had we taken $G_1(\Theta)= (1/n)\sin n\Theta$ and $n\in \mathbb{N}$, the equation of the line $AB$ in Fig.~\ref{Figid} would be $\Theta=2\pi/n$ to ensure the smoothness of the vector field as one crosses the line; that is, $\Theta$ would run from $0$ to $2\pi/n$ and the zeros of $\phi^\Theta=\cos n\Theta=0$ would be at $\pi/(2n)$ and $3\pi/(2n)$, which are both smaller than the upper limit of $\Theta$ ($2\pi/n$). The rules~\eqref{r1} and~\eqref{r2} would take the form
\begin{align}
	&\text{For }G_1(\Theta)=\frac{\sin n\Theta}{n}\,,\nonumber\\
	&\text{If }F(r_h=x)=\text{ max},\quad X_1=(x,\,\pi/(2n)),\quad  X_2=(x,\,3\pi/(2n))\Longrightarrow \nonumber\\
	\label{r1n}&\text{ind}(\phi,X_1)>0 \text{ and }\text{ind}(\phi,X_2)<0\,,\\
	&\text{If }F(r_h=x)=\text{ min},\quad X_1=(x,\,\pi/(2n)),\quad  X_2=(x,\,3\pi/(2n))\Longrightarrow \nonumber\\
	\label{r2n}&\text{ind}(\phi,X_1)<0 \text{ and }\text{ind}(\phi,X_2)>0\,,
\end{align}
and similar rules for $G_2(\Theta)=(1/n)\cos n\Theta$.

From now on we will only consider the function $G_1(\Theta)= (1/2)\sin 2\Theta$ and, for topological classification, we define the topological charge $Q$ of the black hole as the sum of the indices (which is the total winding number $W$) of all CPs on the line $\Theta=3\pi/4$:
\begin{equation}\label{Q}
	Q=\sum _{i}^{(\Theta=3\pi/4)}\text{ind}(\phi,X_i)\,.
\end{equation}

\section{Topological classes\label{secTC}}
As we said earlier we consider the function $G_1(\Theta)=(1/2)\sin 2\Theta$ and the CPs on the horizontal line $\Theta=3\pi/4$ so that the topological charge is given by~\eqref{Q}. There is a couple of configurations to consider in order to determine the topological classes in terms of the CPs. In mathematical analysis a CP of $F(r_h)$ is a point $r_h=x$ where $\partial_{r_h}F(r_h)\big|_{r_h=x}=0$. The real number $F(x)$ is not necessarily a maximum or minimum value [for instance, consider the value of $x^3$ at its unique CP ($x=0$), which is neither a local maximum nor minimum].

\begin{figure}[!htb]
	\centering
	\includegraphics[width=0.6\textwidth]{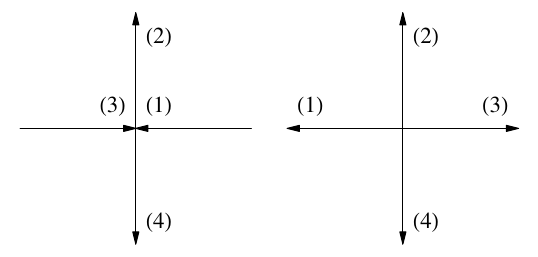}
	\caption{$F(r_h)$ has a maximum at the CP $r_h=x$. The horizontal line is $\Theta=3\pi/4$ and the vertical line is $r_h=x$. In the left panel, the unit vectors are directed from lower to higher values of the corresponding function. In the right panel, the vectors have been translated to have a common tail.}
	\label{Figmax}
\end{figure}
\begin{figure}[!htb]
	\centering
	\includegraphics[width=0.6\textwidth]{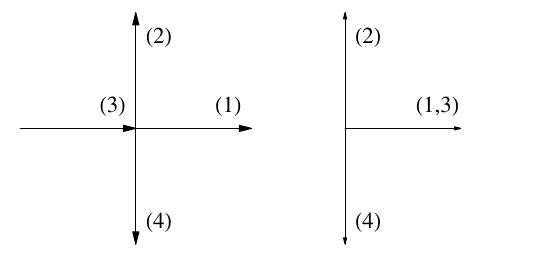}
	\caption{The increasing $F(r_h)$ has neither a maximum nor minimum at the CP $r_h=x$: a point of inflection. The horizontal line is $\Theta=3\pi/4$ and the vertical line is $r_h=x$. In the left panel, the unit vectors are directed from lower to higher values of the corresponding function. In the right panel, the vectors have been translated to have a common tail.}
	\label{Fignomax}
\end{figure}

In the above rules, \eqref{r1} and \eqref{r2}, we focused for simplicity on the case where $F(x)$ is a maximum or a minimum. Before we consider the case where $F(x)$ is neither a local maximum nor minimum, we present a simple method for evaluating the winding number, and we apply it, as an instance, to the case where $F(x)$ is maximum, then apply it to more general cases.

In the left panel of Fig.~\ref{Figmax} we consider a function $F(r_h)$ that has a maximum at the CP $r_h=x$. In line $\Theta=3\pi/4$, the unit vectors (1) and (3) are directed from lower to higher values of $F(r_h)$. The unit vectors (2) and (4) are directed from lower values to higher values of $G_1(\Theta)$. In the right panel, we have translated the four vectors so that they have the same common tail. It is easy to see that the rotation in the order $(1)\to(2)\to(3)\to(4)\to(1)$ \emph{along the shortest path} occurs in the clockwise direction and produces a total rotation angle of $2\pi$ and a winding number of $-1$. 

In the left panel of Fig.~\ref{Fignomax} we consider an increasing function $F(r_h)$ that has neither a maximum nor a minimum at the CP $r_h=x$. This is a point of inflection. In line $\Theta=3\pi/4$, the unit vectors (1) and (3) are directed from lower to higher values of $F(r_h)$. The unit vectors (2) and (4) are directed from lower values to higher values of $G_1(\Theta)$. In the right panel, we have translated the four vectors so that they have the same common tail. It is easy to see that the rotation in the order $(1)\to(2)\to(3)\to(4)\to(1)$ \emph{along the shortest path} yields a total rotation angle of $0$ and a winding number of $0$. From now on, we will not need to include such CPs [where $F(x)$ is neither a maximum or a minimum] in our consideration.

Let $N_{\downarrow}$ and $N_{\uparrow}$ be the number of minima and maxima of $F(r_h)$ as $r_h$ runs from its minimum value $r_m$ to $\infty$, respectively. Then
\begin{align}
	&\text{If }N_{\downarrow}-N_{\uparrow}=0& &\Longrightarrow & & Q=0\,,\nonumber\\
	\label{r3}&\text{If }N_{\downarrow}-N_{\uparrow}=1& &\Longrightarrow & &Q=1\,,\\
	&\text{If }N_{\downarrow}-N_{\uparrow}=-1& &\Longrightarrow & &Q=-1\,.\nonumber
\end{align}
Recall that we are considering a smooth function $F(r_h)$, with no jump discontinuities, which has a continuous and smooth derivative $\phi^{r_h}$ at the CP $r_h=x$. This justifies why $|N_{\downarrow}-N_{\uparrow}|\leq 1$.

The rules above~\eqref{r3} can be refined or divided into subcases if $F(r_h)$ is endowed with further properties. In black-hole thermodynamics, we will be able, using the first law along with the Hawking temperature formula, to refine them.

\section{Thermodynamic topological classes\label{secTTC}}
In black-hole thermodynamic topology, $F(r_h)$ is taken as the off-shell free energy, which depends on an additional parameter $\tau$ the inverse of which is the temperature of the cavity that surrounds the black hole~\cite{York}
\begin{equation}\label{t1}
	F(r_h)=M(r_h)-\frac{S(r_h)}{\tau}\,,
\end{equation}
where $M$ and $S$ are mass and entropy of the black hole. In Ref.~\cite{WLM}, on using the first law of black-hole thermodynamics along with Hawking temperature, it was shown that
\begin{equation}\label{t2}
	\phi^{r_h}=\partial_{r_h}S (T-\tau^{-1})\,,
\end{equation}
and it was argued that \emph{in general} the temperature of the black hole at spatial infinity, $T(\infty)$, and at $r_m$ (the minimum value of $r_h$), $T(r_m)$, is either $0$ or $\infty$. This yields four combinatory cases
\begin{align}
	&\text{I: }& &\hspace{-30mm}T(r_m)=\infty,\quad T(\infty)=0\,,\nonumber\\
	\label{r4}	&\text{II: }& &\hspace{-30mm}T(r_m)=0,\quad T(\infty)=0\,,\\
	&\text{III: }& &\hspace{-30mm}T(r_m)=0,\quad T(\infty)=\infty\,,\nonumber\\
	&\text{IV: }& &\hspace{-30mm}T(r_m)=\infty,\quad T(\infty)=\infty\,.\nonumber
\end{align}

\paragraph{Case I:} We assume that $\partial_{r_h}S>0$. On the line $\Theta=3\pi/4$, we start moving from $r_m$ until we reach the first CP (from left to right), $T$ decreases from $\infty$, becomes equal to $\tau^{-1}$, and then smaller than $\tau^{-1}$, so that $T-\tau^{-1}$ changes sign from plus to minus. This corresponds to a maximum of $F(r_h)$ and a relative winding number of $-1$. Similarly, we start moving from large $r_h$ (from right to left) until we reach the last CP, $T$ increases from $0$, becomes equal to $\tau^{-1}$, and then greater than $\tau^{-1}$, so that $T-\tau^{-1}$ changes sign from minus to plus. This again corresponds to a maximum of $F(r_h)$ and a relative winding number of $-1$. Case I corresponds to $N_{\downarrow}-N_{\uparrow}=-1$~\eqref{r3} and a total winding number $Q=-1$. This is represented symbolically as $Q^{1-}=[\uparrow,\uparrow]$ to say that the first and last CPs (as $r_h$ runs from $r_m$ to $\infty$ on the line $\Theta=3\pi/4$) correspond to maxima of $F(r_h)$ and the total winding number is $-1$. The notation $W^{1-}=[-,-]$, used in~\cite{WLM}, is used to say that the first and last CPs have relative winding numbers of $-1$.

\paragraph{Case II:} We assume that $\partial_{r_h}S>0$. On the line $\Theta=3\pi/4$, we start moving from $r_m$ until we reach the first CP (from left to right), $T$ increases from $0$, becomes equal to $\tau^{-1}$, and then greater than $\tau^{-1}$, so that $T-\tau^{-1}$ changes sign from minus to plus. This corresponds to a minimum of $F(r_h)$ and a relative winding number of $+1$. Similarly, we start moving from large $r_h$ (from right to left) until we reach the last CP, $T$ increases from $0$, becomes equal to $\tau^{-1}$, and then greater than $\tau^{-1}$, so that $T-\tau^{-1}$ changes sign from minus to plus. This corresponds to a maximum of $F(r_h)$ and a relative winding number of $-1$. Case II corresponds to $N_{\downarrow}-N_{\uparrow}=0$~\eqref{r3} and a total winding number $Q=0$. This is represented symbolically as $Q^{0+}=[\downarrow,\uparrow]$ to say that the first CP and the last CP (as $r_h$ runs from $r_m$ to $\infty$ on the line $\Theta=3\pi/4$) correspond to a minimum and maximum of $F(r_h)$, respectively, and that the total winding number is $0$. The notation $W^{0+}=[+,-]$, used in~\cite{WLM}, is used to say that the first CP and the last CP have relative winding numbers of $+1$ and $-1$, respectively.

\paragraph{Case III:} On repeating the same arguments as in Cases I and II, we conclude that Case III corresponds to $N_{\downarrow}-N_{\uparrow}=1$~\eqref{r3} and a total winding number $Q=1$. This is represented symbolically as $Q^{1+}=[\downarrow,\downarrow]$ to say that the first and last CPs (as $r_h$ runs from $r_m$ to $\infty$ on the line $\Theta=3\pi/4$) correspond to the minima of $F(r_h)$ and the total winding number is $1$. The notation $W^{1+}=[+,+]$, used in~\cite{WLM}, is used to say that the first and last CPs have relative winding numbers of $1$.

\paragraph{Case IV:} Case IV corresponds again to $N_{\downarrow}-N_{\uparrow}=0$~\eqref{r3} and to a total winding number $Q=0$. This is represented symbolically as $Q^{0-}=[\uparrow,\downarrow]$ to say that the first CP and the last CP (as $r_h$ runs from $r_m$ to $\infty$ on the line $\Theta=3\pi/4$) correspond to a maximum and minimum of $F(r_h)$, respectively, and that the total winding number is $0$. The notation $W^{0-}=[-,+]$, used in~\cite{WLM}, is used to say that the first CP and last CP have relative winding numbers of $-1$ and $+1$, respectively.

\section{Application\label{secAp}}

\subsection{Lorentzian Taub-NUT Black Hole}
The line element of the Lorentzian Taub-NUT black hole can be written as \cite{newman1963empty}
\begin{equation}\label{LTNUT}
	ds^2=-\frac{f(r)}{r^2+n^2}\left(dt+2n\cos\theta d\phi\right)^2+\frac{r^2+n^2}{f(r)}dr^2+\left(r^2+n^2\right)\left(d\theta^2+\sin^2\theta d\phi^2\right),
\end{equation}
with the function 
\begin{equation}
	f\left(r\right)=r^2-2mr-n^2,
\end{equation}
and $m$ and $n$ being the mass and NUT charge parameters, respectively. Here, the NUT-charge parameter $n$ represents a gravitomagnetic charge and is responsible for the twist of spacetime. The Lorentzian Taub-NUT black hole, similar to the Schwarzschild black hole, has only one horizon, the event horizon, which is the positive root of the horizon equation $f(r_h) = 0$ and is located at $r_h = m + \sqrt{m^2 + n^2}$. Note that the size of the event horizon increases with increasing the NUT-charged parameter $n$. 

The black hole mass $M$ and other thermodynamical quantities required for topological classification are given as \cite{hennigar2019thermodynamics}
\begin{eqnarray}
	M=m, \quad S=\pi \left(r_h^2+n^2\right), \quad  \mathcal{N}=-\frac{4\pi n^3}{r_h}, \quad \mathcal{\psi}= \frac{1}{8\pi n},\quad T= \frac{1}{4\pi r_h}\,,
\end{eqnarray}
where $S$ and $T$ represent the entropy and Hawking temperature, whereas $\mathcal{N}$ represents the gravitational Misner charge and $\mathcal{\psi}$ the corresponding associated potential. We see the that temperature $T$ has the following limits
\begin{eqnarray}
	T(r_m)=\infty \quad \text{and} \quad T(\infty)=0,
\end{eqnarray}
where $r_m=0$ for the Lorentzian Taub-NUT black hole spacetime. By~\eqref{r4}, this has the thermodynamic properties of Case I: $Q^{1-}=[\uparrow,\uparrow]=[-,-]$.

The on-shell free energy in this case can be written as 
\begin{eqnarray}
	\mathcal{F}=M-ST-\mathcal{\psi}\mathcal{N}. 
\end{eqnarray}
Thus, the from the modified off-shell free energy takes the form
\begin{eqnarray}\label{FpG}
	\tilde{\mathcal{F}}=F(r_h)+G_1(\Theta)=\frac{r_h}{2}-\frac{\pi \left(r_h^2+n^2\right)}{\tau}+\frac{\sin 2\Theta}{2},
\end{eqnarray}
where we took $G=G_1=(1/2)\sin 2\Theta$, and the $\phi$-vector field is the gradient of $\tilde{\mathcal{F}}$ and obeys the general formula~\eqref{F2}
\begin{eqnarray}\label{LTNUT2}
	\phi^{r_h}=\frac{1}{2}-\frac{2\pi r_h }{\tau}, \quad \phi^{\Theta}=\cos 2\Theta\,.
\end{eqnarray}
We see from~\eqref{FpG} that the presence of the ``$-$'' sign indicates that $F(r_h)$ has a maximum at the unique zero of $\phi^{r_h}=0$, which is $x=\tau/(4\pi)$. By the rule~\eqref{r1}, on setting $X_1=(\tau/(4\pi),\,\pi/4)$ and $X_2=(\tau/(4\pi),\,3\pi/4)$, we have $\text{ind}(\phi,X_1)>0$ and $\text{ind}(\phi,X_2)<0$. Vector field plots of the unit vector field $\phi/||\phi||$ is depicted in the left panel of Fig.~\ref{FigTNUT} (in the right panel of this figure we depict the same plot taking $G=G_2=(1/2)\cos 2\Theta$). We see from the figure that the total winding number (topological charge) is $Q=W=-1$.

%

\subsection{Lorentzian Reissner-Nordstr\"{o}m-NUT Black Hole}
The line element is given by
\begin{multline}\label{rnmetric}
	ds^2=-\frac{f(r)}{r^2+n^2}(dt+2n\cos\theta\,d\varphi)^2 + \frac{r^2+n^2}{f(r)}~dr^2
	+(r^2+n^2)(d\theta^2+\sin^2\theta\,d\varphi^2)\,,
\end{multline}
where $f(r)=r^2-2mr-n^2+q^2$. The horizon is $r_h=m+\sqrt{m^2+n^2-q^2}$ and the horizon of the extreme black hole is $r_h=r_m=m$. The Hawking temperature is calculated by
\begin{equation}\label{rn1}
	T=\frac{f'(r_h)}{4\pi(r_h^2+n^2)}=\frac{r_h-m}{2\pi(r_h^2+n^2)}\,,
\end{equation}
yielding $T(r_m)=0$. The value of $T(\infty)$ is evaluated first by eliminating $m$ from the expression of $T$ using $f(r_h)=0$: $m=(r_h^2+q^2-n^2)/(2r_h)$. We obtain
\begin{equation}\label{rn2}
	T=\frac{r_h^2+n^2-q^2}{4\pi r_h(r_h^2+n^2)}\,,
\end{equation}
and $T(\infty)=0$. By~\eqref{r4}, this has the thermodynamic properties of Case II: $Q^{0+}=[\downarrow,\uparrow]=[+,-]$.

\subsection{Lorentzian Kerr-NUT Black Hole}
The line element of the Lorentzian Kerr-NUT black hole 
\begin{multline}\label{knut}
	ds^2=-\frac{\Delta}{\Sigma}\left[dt+(2n\cos\theta - a \sin^2\theta)d\phi\right]^2+\frac{\Sigma}{\Delta}dr^2+\Sigma d\theta^2
	+\frac{\sin^2\theta}{\Sigma}\left[adt-(r^2+a^2+n^2 d\varphi\right]^2,
\end{multline}
with the functions
\begin{eqnarray}
	\Delta=r^2-2mr+a^2-n^2, \quad \text{and} \quad \Sigma=r^2+(n+a\cos\theta)^2. 
\end{eqnarray}
and $a$ being a rotation parameter. The other entities are, respectively, the mass parameter $m$ and the NUT charge $n$. The black hole horizon is located at the largest root of $\Delta(r_h)=0$: $r_h=m+\sqrt{m^2+n^2-a^2}$. The extreme black hole has $r_h=r_m=m$.

Let $g(r)=r^2-2mr-n^2$. The Hawking temperature is calculated by
\begin{equation}\label{kn1}
	T=\frac{g'(r_h)}{4\pi(r_h^2+a^2+n^2)}=\frac{r_h-m}{2\pi(r_h^2+a^2+n^2)}\,.
\end{equation}
We immediately obtain $T(r_m)=0$. Using the fact that the black hole horizon is located at the largest root of $\Delta(r_h)=0$, yielding $m=(r_h^2+a^2-n^2)/(2r_h)$ and
\begin{equation}\label{kn2}
	T=\frac{r_h^2+n^2-a^2}{4\pi r_h(r_h^2+n^2+a^2)}\,,
\end{equation}
from which we obtain $T(\infty)=0$. By~\eqref{r4}, this too has the thermodynamic properties of II: $Q^{0+}=[\downarrow,\uparrow]=[+,-]$.

\subsection{Lorentzian Kerr-Newman-NUT Black Hole}
The line element is given by~\eqref{knut} with $\Delta=r^2-2mr+a^2+q^2-n^2$. The horizon is located at $r_h=m+\sqrt{m^2+n^2-a^2-q^2}$. The extreme black hole has $r_h=r_m=m$. With the black hole temperature given by~\cite{Yang}
\begin{equation}\label{T1}
	T=\frac{r_h-m}{2\pi(r_h^2+a^2+n^2)}=\frac{r_h^2+n^2-a^2-q^2}{4\pi r_h(r_h^2+n^2+a^2)}\,,
\end{equation}
we see that $T(r_m)=0$ and $T(\infty)=\infty$. So, by~\eqref{r4}, this too has the thermodynamic properties of II: $Q^{0+}=[\downarrow,\uparrow]=[+,-]$.

\subsection{Lorentzian Taub-NUT-AdS Black Hole}
The metric is given by~\eqref{rnmetric} with $f(r)=r^2-2mr-n^2+(r^4+6n^2r^2-3n^4)/\ell^2$.

For this black hole we will determine the extreme values of $F(r_h)$, which is given by~\cite{wu2023classifying}
\begin{equation}\label{tna1}
	F(r_h)=\frac{r_h}{2}+\frac{4\pi Pr_h(r_h^2+3n^2)}{3}-\frac{\pi (r_h^2+n^2)}{\tau}\,,\qquad P=\frac{3}{8\pi \ell^2}=-\frac{\Lambda}{8\pi}\,,
\end{equation}
\begin{equation}\label{tna2a}
	\phi^{r_h}=\frac{1}{2}+4\pi P(r_h^2+n^2)-\frac{2\pi r_h}{\tau}\,.
\end{equation}
If $\tau^2 >\pi/(2P+16\pi n^2P^2)$,  $\phi^{r_h}$ never vanishes; that is, there is no black hole solution, since $F(r_h)$ has no extreme values. If $\tau^2 <\pi/(2P+16\pi n^2P^2)$, there is a small black hole at $r_{h-}$ and a large black hole at black hole at $r_{h+}$
\begin{equation}\label{tna3}
	r_{h\pm}=\frac{1\pm \sqrt{1-16n^2P^2\tau^2-2P\tau^2/\pi}}{4P \tau}	\,.
\end{equation}
$F(r_{h-})$ is a maximum value and $F(r_{h+})$ is a minimum value, so the total winding number is $0$. By~\eqref{r4}, this has the thermodynamic properties of Case IV: $Q^{0-}=[\uparrow,\downarrow]=[-,+]$.

If $\tau^2 =\pi/(2P+16\pi n^2P^2)$, there is one root to $\phi^{r_h}=0$ with $r_h=1/(4P\tau)$. This is a CP but $F(r_h)$ is not an extreme value, so by Fig.~\ref{Fignomax} the total winding number is again $0$.

We reach the same conclusion using the expression of temperature
\begin{equation}\label{T}
	T=\frac{1}{4\pi r_h}\Big(1+3~\frac{r_h^2+n^2}{\ell^2}\Big)\,.
\end{equation}
This black hole admits no extreme case, and thus $r_m=0$. Consequently, we have $T(r_m)=\infty$ and $T(\infty)=\infty$. This is again Case IV: $Q^{0-}=[\uparrow,\downarrow]=[-,+]$.

\subsection{Lorentzian Reissner-Nordstr\"{o}m-NUT-AdS Black Hole}
The metric is given by~\eqref{rnmetric} with $f(r)=r^2-2mr-n^2+q^2+(r^4+6n^2r^2-3n^4)/\ell^2$.

We will proceed as in the previous section. The $F(r_h)$ is given by~\cite{wu2023consistent}
\begin{equation}\label{l1}
	F(r_h)=\frac{r_h}{2}+\frac{4\pi Pr_h(r_h^2+3n^2)}{3}-\frac{\pi (r_h^2+n^2)}{\tau}+\frac{q^2r_h(r_h^2-n^2)}{2(r_h^2+n^2)^2}\,,\qquad P=\frac{3}{8\pi \ell^2}=-\frac{\Lambda}{8\pi}\,,
\end{equation}
and
\begin{equation}\label{l2}
	\phi^{r_h}=\frac{1}{2}+4\pi P(r_h^2+n^2)-\frac{2\pi r_h}{\tau}-\frac{q^2(r_h^4-6n^2r_h^2+n^4)}{2(r_h^2+n^2)^3}\,.
\end{equation}

The temperature is given by~\cite{wu2023consistent}
\begin{equation}\label{l3}
	T=\frac{1}{4\pi r_h}\Big(1-\frac{q^2}{r_h^2+n^2}+3~\frac{r_h^2+n^2}{\ell^2}\Big)\,,
\end{equation}
yielding $T(\infty)=\infty$. By~\eqref{r4}, this has the thermodynamic properties of Case III or Case IV. This depends on the choice of parameters.

For
\begin{equation}\label{cond1}
	q^2<n^2+\frac{3n^4}{\ell^2}\,,
\end{equation}
the black hole admits no extreme case, and thus $r_m=0$. Consequently, we have $T(r_m)=\infty$ and $T(\infty)=\infty$. By~\eqref{r4}, this has the thermodynamic properties of Case IV: $Q^{0-}=[\uparrow,\downarrow]=[-,+]$. This allows us to generalize the result of the previous section: If
\begin{equation}\label{cond11}
	0\leq q^2<q_c^2 \equiv n^2+\frac{3n^4}{\ell^2}=n^2+8\pi Pn^4\,,
\end{equation}
the black hole has the thermodynamic properties of Case IV: $Q^{0-}=[\uparrow,\downarrow]=[-,+]$.

For
\begin{equation}\label{cond2}
	q^2\geq q_c^2 \equiv n^2+\frac{3n^4}{\ell^2}=n^2+8\pi Pn^4\,,
\end{equation}
the black hole has an extreme case with horizon given by
\begin{equation}\label{cond21}
	r_h=r_m=\frac{\sqrt{\ell\sqrt{12q^2+\ell^2}-\ell^2-6n^2}}{\sqrt{6}}\geq 0\,,
\end{equation}
with $T(r_m)=0$. By~\eqref{r4}, this has the thermodynamic properties of Case III: $Q^{1+}=[\downarrow,\downarrow]=[+,+]$. Note that $r_m=0$ if the equality in \eqref{cond2} holds; in this case $T=r(3r^2+6n^2+\ell^2)/[4\pi\ell^2(r^2+n^2)]$, and so $T(r_m)=0$. 

The extreme black hole has the following extremity constraint
\begin{multline}\label{const}
	m^2=\frac{2}{9} (2 q^2-5 n^2)-\frac{16 n^6}{\ell ^4}+\frac{8 n^2 (q^2-3 n^2)}{3 \ell ^2}-\frac{\ell ^2}{27}\\+\frac{1}{r_m^2}~\Big[\frac{1}{9} (4 q^4-4
	n^2 q^2-n^4)-\frac{4 n^4 q^2}{3 \ell ^2}+\frac{1}{27} (q^2-n^2) \ell ^2\Big]\,,
\end{multline}
where $r_m$ is given in~\eqref{cond21}. This reduces to $m^2=q^2$ if $n=0$ and $\ell\to\infty$, and reduces to $m^2=q^2-n^2=r_h^2$ if only $\ell\to\infty$. If the equality in~\eqref{cond2} is satisfied, Eq.~\eqref{const} reduces to $m^2=(6n^2+\ell^2)^3/(27\ell^4)$. To our knowledge, the expressions \eqref{cond2}, \eqref{cond21} and \eqref{const} are not available in the scientific literature. These expressions could also have been derived by analyzing the roots of $f(r_h)=0$.

There is a thermodynamic topological transition $Q^{0-}\to Q^{1+}$ as the electric charge changes values from $q^2<q_c^2=n^2+8\pi Pn^4$ to $q^2\geq q_c^2$. To see the transition clearly, we plot $\tau$ versus $r_h$, where $\tau(r_h)$ is obtained from~\eqref{l2} on setting $\phi^{r_h}=0$~\cite{wu2023consistent}
\begin{equation}\label{tr1}
	\tau =\frac{4 \pi  r_h (n^2+r_h^2)^3}{8 \pi P(r_h^2+n^2)^4+(r_h^2+n^2)^3-q^2 (r_h^4-6 n^2 r_h^2+n^4)}\,.
\end{equation}
First of all, we note that the denominator of this expression reduces to $8 \pi  P r_h^8+(1+32 n^2 \pi  P) r_h^6+2 n^2 (1+20 n^2 \pi  P) r_h^4+n^4 (9+80 n^2 \pi  P) r_h^2$ if $q^2= q_c^2$, which means that the graph ($r_h,\,\tau(r_h)$) will have a vertical asymptote as $r_h\to 0$ at the transition point, that is, $\tau(r_h)\to\infty$ as $r_h\to 0$ if $q^2= q_c^2$ [but $\tau(r_h=0)=0$ if $q^2\neq q_c^2$]. This is quit different from the transition $Q^{0-}\to Q^{1+}$, first discovered in~\cite{trans1} and further discussed in~\cite{trans2}, where $\tau(r_h)\to\text{ finite value }>0$ as $r_h\to 0$ if $q^2= q_c^2$ (here $q_c^2$ is some critical value expressed in terms of the BH parameters of~\cite{trans1}).

\begin{figure}[!htb]
	\centering
	\includegraphics[width=0.4963\textwidth]{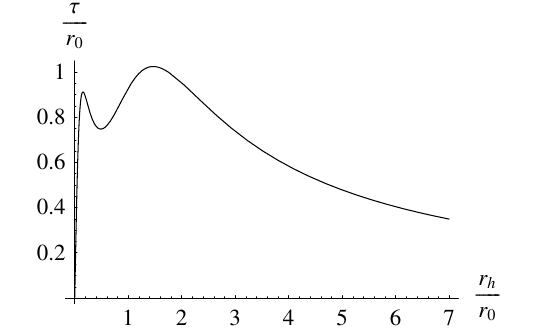}
	\includegraphics[width=0.4963\textwidth]{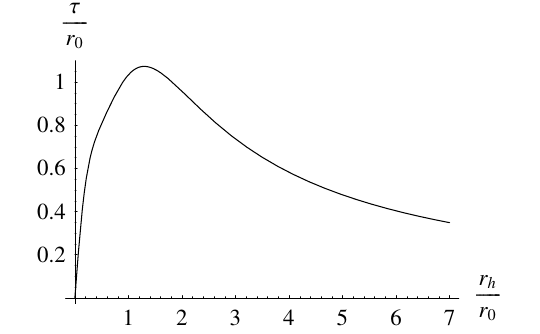}
	\caption{Roots of $\phi^{r_h}=0$ in the $r_h\tau$-plane for $q^2<q_c^2=n^2+8\pi Pn^4$ taking $n=r_0$, $P=2/(10r_0^2)$, $q_c^2=(5+8\pi)/5$. Left Panel: $q^2=(-14+8\pi)/5$. Right Panel: $q^2=(0+8\pi)/5$.}
	\label{FigTrans1}
\end{figure}
\begin{figure}[!htb]
	\centering
	\includegraphics[width=0.4963\textwidth]{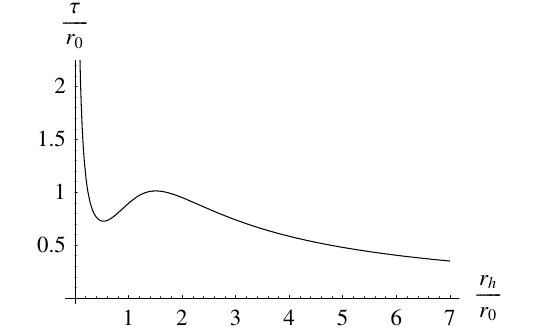}
	\includegraphics[width=0.4963\textwidth]{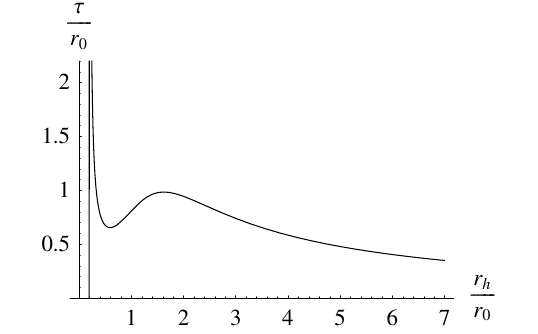}
	\caption{Roots of $\phi^{r_h}=0$ in the $r_h\tau$-plane for $q^2\geq q_c^2=n^2+8\pi Pn^4$ taking $n=r_0$, $P=2/(10r_0^2)$, $q_c^2=(5+8\pi)/5$. Left Panel: $q^2=q_c^2$. Right Panel: $q^2=(20+8\pi)/5$.}
	\label{FigTrans2}
\end{figure}

In Figs.~\ref{FigTrans1} and~\ref{FigTrans2}, a section of the graph where $\tau(r_h)$ increases as $r_h$ does, corresponds to an unstable black hole, and a section of the graph where $\tau(r_h)$ decreases as $r_h$ increases, corresponds to a stable black hole. 

In Fig.~\ref{FigTrans1} we have $q^2<q_c^2$, any horizontal line $\tau=\text{const.}$ intersects the plot at an even number of points, so that $N_{\downarrow}-N_{\uparrow}=0$~\eqref{r3}. Since the leftmost point corresponds to an unstable black hole, this is the class $Q^{0-}$. If $\tau_{\text{ext}}$ is an extreme value of $\tau(r_h)$, the horizontal line $\tau=\tau_{\text{ext}}$ intersects the plot at an odd number of points, however, the root $r_h$ corresponding to $\tau=\tau_{\text{ext}}$ is a point of inflection of $F(r_h)$ and the corresponding winding number is $0$ by Fig.~\ref{Fignomax}, and still $Q=0$. There is no vertical asymptote at $r_h=0$ and we have $\tau(0)=0$.

In Fig.~\ref{FigTrans2} we have $q^2\geq q_c^2$, any horizontal line $\tau=\text{const.}$ intersects the plot at an odd number of points, so that $N_{\downarrow}-N_{\uparrow}=1$~\eqref{r3}. Since the leftmost point corresponds to a stable black hole, this is the class $Q^{1+}$. If $\tau_{\text{ext}}$ is an extreme value of $\tau(r_h)$, the horizontal line $\tau=\tau_{\text{ext}}$ intersects the plot at an even number of points, however, the root $r_h$ corresponding to $\tau=\tau_{\text{ext}}$ is a point of inflection of $F(r_h)$ and the corresponding winding number is $0$ by Fig.~\ref{Fignomax}, and still $Q=1$. There is a vertical asymptote at $r_h=0$ if $q^2= q_c^2$, and for $q^2> q_c^2$ this vertical asymptote is shifted rightward. It is worth mentioning that vertical asymptotes are not part of the graph ($r_h,\,\tau(r_h)$), and an intersection of a horizontal line with a vertical asymptote does not correspond to an extreme value of $F(r_h)$, consequently such an intersection does not change any of the numbers $N_{\downarrow}$ and $N_{\uparrow}$.

Upon comparing the plots in Figs.~\ref{FigTrans1} and~\ref{FigTrans2}, we see that the transition $Q^{0-}\to Q^{1+}$ occurs when a small black hole ($r_h\to 0$), with $q^2<q_c^2$, absorbs more electric charge from its environment, converting the unstable section of the plot near $r_h=0$ to a stable one as soon as $q^2$ becomes equal to $q_c^2$. This state of stability persists when $q^2$ exceeds $q_c^2$. The transition $Q^{0-}\to Q^{1+}$ discussed here is charge-dependent, while the same transition $Q^{0-}\to Q^{1+}$ discovered in~\cite{trans1} is rather temperature-dependent, that is, for fixed $q^2<q_c^2$ (here $q_c^2$ is some critical value expressed in terms of the BH parameters of~\cite{trans1}), the transition occurs as the temperature of the black hole changes, while for $q^2\geq q_c^2$ no such temperature-dependent transition occurs.
 
\section{Conclusion\label{secConc}}
We have developed new analytical methods that allow for the topological selections and classifications of vector fields. Taking advantage of the Poincar\'e-Hopf Index Theorem for manifold without boundary, the methods consist of 1) dropping the usually employed vector component function $\csc$ and replacing it by analytical expressions that include the $\cos$ or $\sin$ function; 2) obtaining simple rules for the determination of the index of a vector field; 3) relating the total topological charge of a vector field to the difference of the number of maxima and minima of the potential function from which the vector field is derived. As a first application, we rederived the four cases of thermodynamic topological classes.

As a second application, we considered a set of black hole solutions and determined their classes. We summarize our results, obtained in the last section, as follows.
\begin{align}
	&\text{Taub-NUT BH}& &\text{Case I: }Q^{1-}=[\uparrow,\uparrow]=[-,-]\nonumber\\
	&\text{Reissner-Nordstr\"{o}m-NUT BH}& &\text{Case II: } Q^{0+}=[\downarrow,\uparrow]=[+,-]\nonumber\\
	&\text{Kerr-NUT BH}& &\text{Case II: } Q^{0+}=[\downarrow,\uparrow]=[+,-]\nonumber\\
	&\text{Kerr-Newman-NUT BH}& &\text{Case II: } Q^{0+}=[\downarrow,\uparrow]=[+,-]\nonumber\\
	&\text{Taub-NUT-AdS BH}& &\text{Case IV: } Q^{0-}=[\uparrow,\downarrow]=[-,+]\nonumber\\
	&\text{Reissner-Nordstr\"{o}m-NUT-AdS BH}& &\begin{cases}
		q^2<n^2+\frac{3n^4}{\ell^2}& \text{Case IV: } Q^{0-}=[\uparrow,\downarrow]=[-,+]\\
		q^2\geq n^2+\frac{3n^4}{\ell^2}& \text{Case III: } Q^{1+}=[\downarrow,\downarrow]=[+,+]
	\end{cases}
	\nonumber
\end{align}
We see that all Taub-NUT charged and/or rotating black holes are in Class II; neutral and non-rotating Taub-NUT black holes are in Class I. AdS black holes are in Class IV if their electric charge squared is zero or smaller than $n^2+3n^4/\ell^2$; otherwise, if $q^2\geq n^2+3n^4/\ell^2$, the AdS black holes will be in Class III. For the Reissner-Nordstr\"{o}m-NUT-AdS BH, we noticed a charge-dependent thermodynamic topological transition $Q^{0-}\to Q^{1+}$.

For the Reissner-Nordstr\"{o}m-NUT-AdS black hole, a numerical example was given in~\cite{wu2023consistent} taking $q=r_0$, $n=r_0$, $P=0.2/r_0^2$ ($r_0$ is an arbitrary length scale) and this yielded $Q=0$ with two positive roots to $\phi^{r_h}=0$ (one maximum and one minimum), which corresponds to Case IV. We have checked that for $q=r_0$, $n=r_0$, and $P=0.0002/r_0^2$, there are four positive roots to $\phi^{r_h}=0$ (two maxima and two minima) and this corresponds to Case IV too. To provide a numerical example supporting Case III, the electric charge has to be relatively high: For $q=5.5r_0$, $n=r_0$, and $P=0.0002/r_0^2$, there are three positive roots to $\phi^{r_h}=0$ (two minima and one maximum).

\appendix
\section{The Euler characteristic of a cylinder}
Subdivisions or triangulation is the simplest way to calculate the Euler characteristic $\chi(\mathcal{M})$ of a surface $\mathcal{M}$. The formula reads
\begin{equation}\label{a1}
	\chi(\mathcal{M})=V-E+F\,,
\end{equation}
where $V$ is the number of vertices in the subdivision, $E$ the number of edges, and $F$ the number of faces. It is not necessary to divide the surface into a collection of finitely many triangles as done in~\cite{book2}; subdivisions into other shapes, such as squares, offer much simpler ways to calculate the Euler characteristic. 

In general, a subdivision of a surface consists of a finite set of vertices, and a finite set of edges such that 1) each vertex is an endpoint of at least one edge, 2) the vertices and edges form a connected graph, 3) the space enclosed by the edges is a face, 4) faces are disjoint pieces, each of which is homeomorphic to an open disk.

The construction shown in the right panel of Fig.~\ref{Figid} is called paper-and-glue construction, which consists of making a rectangle into a cylinder by gluing together two opposite edges. To calculate the Euler characteristic, we apply the paper-and-cut construction, which is the reverse operation of the paper-and-glue construction. From the rectangle shown in the right panel of Fig.~\ref{Figid}, we remove the lower edge $A'\to B'$ (corresponding to $\Theta=0$) along with its vertices $A'$ and $B'$. There are still two vertices $A$ and $B$ and three edges, so $V=2$ and $E=3$. Since we have only one face, $F=1$, resulting in $\chi(\mathcal{M})=2-3+1=0$. Any other subdivision of the rectangle (after deleting the edge $A'\to B'$) into smaller squares or triangles, will results in the same value $\chi(\mathcal{M})=0$.

\newpage

\bibliographystyle{JHEP}
\bibliography{biblio.bib}

\end{document}